\documentclass[preprint, amsmath, amssymb, aps, prb]{revtex4-1}

\usepackage[T1]{fontenc}
\usepackage{graphicx}
\usepackage{hyperref}
\usepackage[version=4]{mhchem}
\usepackage{amsmath}
\usepackage{placeins}

\usepackage[resetlabels]{multibib}
\newcites{SI}{References}

\begin{document}

\title{Ferrimagnetism induced by thermal vibrations in oxygen-deficient manganite heterostructures}

\author{Moloud Kaviani}
\affiliation{Department of Chemistry, Biochemistry and Pharmaceutical Sciences, University of Bern, Freiestrasse 3, 3012 Bern, Switzerland}

\author{Chiara Ricca}
\affiliation{Department of Chemistry, Biochemistry and Pharmaceutical Sciences, University of Bern, Freiestrasse 3, 3012 Bern, Switzerland}

\author{Ulrich Aschauer}
\affiliation{Department of Chemistry, Biochemistry and Pharmaceutical Sciences, University of Bern, Freiestrasse 3, 3012 Bern, Switzerland}
\affiliation{Department of Chemistry and Physics of Materials, University of Salzburg, Jakob-Haringer-Strasse 2a, 5020 Salzburg, Austria}
\email{ulrich.aschauer@plus.ac.at}

\date{\today}
 
\begin{abstract}
Super-exchange most often leads to antiferromagnetim in transition-metal perovskite oxides, yet ferromagnetism or ferrimagnetism would be preferred for many applications, for example in data storage. While alloying, epitaxial strain and defects were shown to lead to ferromagnetism, engineering this magnetic order remains a challenge. We propose, based on density functional theory calculations, a novel route to defect-engineer ferrimagnetism, which is based on preferential displacements of oxygen vacancies due to finite temperature vibrations. This mechanism has an unusual temperature dependence, as it is absent at 0K, strengthens with increasing temperature before vanishing once oxygen vacancies disorder, giving it a unique experimentally detectable signature.
\end{abstract}

\maketitle

Perovskite oxides with the general formula \ce{ABO3} are materials that can flexibly accommodate many elements on the cationic A- and B-sites~\cite{Bhalla2000perovskite}, leading to a large tunability of their properties and consequently applications in many fields such as data storage, optoelectronics or catalysis~\cite{Schlom2008thin, Jeong2012emerging, Zhu2014perovskite}. A notable property of some compositions is the magnetic order they assume due to exchange interactions, at application-relevant temperatures typically between the B-site elements~\cite{Terakura2007magnetism}. While in perovskite oxides antiferromagnetic (AFM) coupling typically dominates between B-sites with the same \textit{d}-electron occupation, doping or alloying on the A-site can lead to a ferromagnetic (FM) order for example in \ce{La_{1-x}Sr_xMnO3}~\cite{Goodenough1955theory, Goodenough1958interpretation, Kanamori1959superexchange}. FM can also emerge due to strain and quantum confinement in ultra-thin films~\cite{Pedroso:2020hd}. In addition to their magnetic properties, some perovskite oxides can be engineered via epitaxial strain to exhibit ferroelectricity thus becoming multiferroic materials~\cite{Spaldin2005renaissance, Ramesh2007multiferroics, Lee2010epitaxial}.

Perovskite manganites (\ce{AMnO3}) are a particularly interesting class of materials due to their proximity to a ferroelectric phase transition~\cite{Marthinsen:2016iq} and the highly flexible oxidation state of the \ce{Mn} B-site. It was shown that free electrons, introduced for example by \ce{La} doping on the \ce{Ca} site in \ce{CaMnO3} are self-trapped in \ce{Mn} $e_g$ states. The exchange interactions favor a FM coupling of the thus reduced \ce{Mn} with its 6 nearest neighbors, forming a 7-site FM ``spin polaron'' in the nominally G-type AFM lattice \cite{Meskine:2004ud, Bondarenko:2019xc}. Oxygen vacancies (\ce{V_O}) are the dominant defect type in perovskite manganites under typical synthesis and application conditions. As opposed to free electrons, the presence of an oxygen vacancy will trap the excess electrons resulting from vacancy formation on $e_g$ states of the two \ce{Mn} adjacent to the vacancy. In an ionic picture, these two sites get reduced from their normal \ce{Mn^{4+}} to a \ce{Mn^{3+}} oxidation state~\cite{Aschauer:2013}. A similar picture is typically observed for perovskite oxides other than manganites as long as they have reducible B-sites, such as ferrites~\cite{Ritzmann:2013kn}. This change in oxidation state (\textit{d}-orbital occupation), sometimes in conjunction with the missing superexchange pathway is known to locally affect the magnetic properties, in extreme cases leading to vacancy-induced FM behavior in AFM materials~\cite{Biskup2014insulating, Gao2018bandgap, Liu2023robust}.

\begin{figure}
	\includegraphics[height=0.8\textheight]{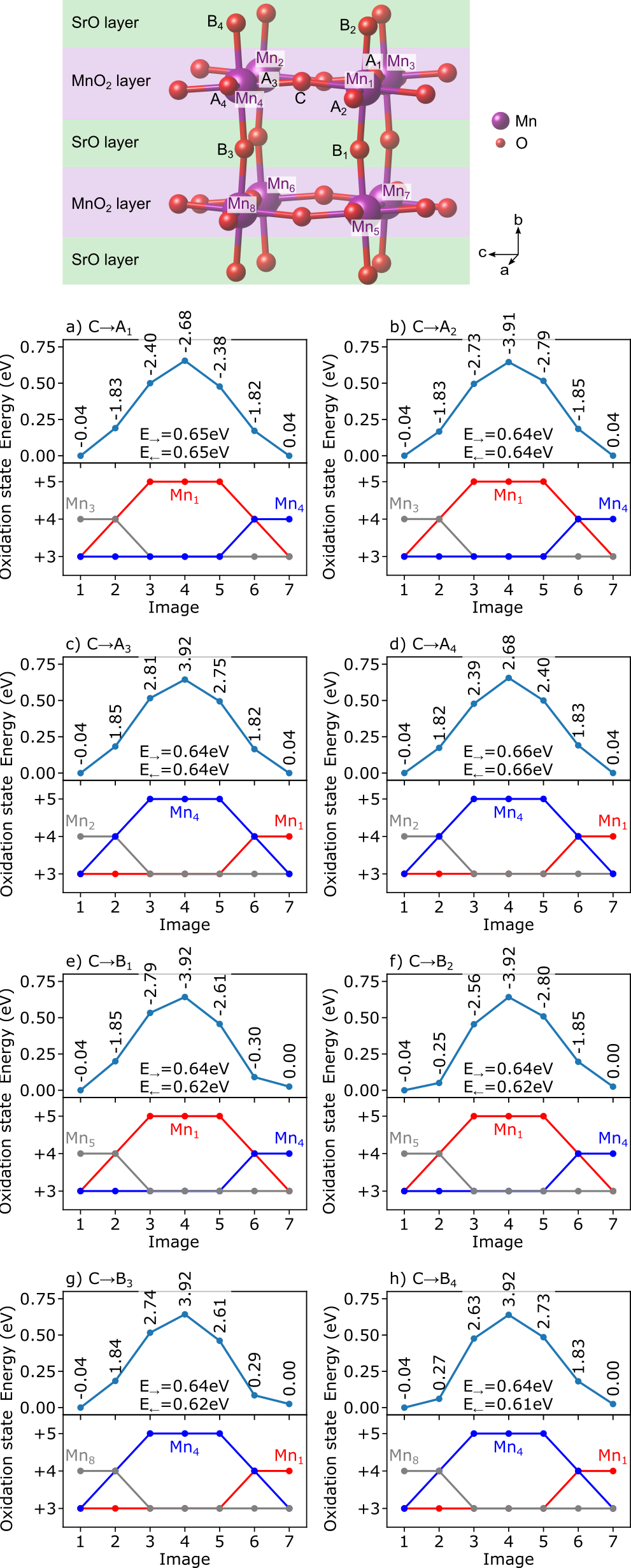}
	\caption{Energy profiles for the migration of a \ce{V_O} in \ce{SrMnO3} from site C to different neighboring sites shown in in the top panel (Sr atoms are omitted for clarity). For each image along the migration paths the total magnetization is reported in $\mu_\textrm{B}$/cell. Barriers for forward (E$_{\rightarrow}$) and reverse (E$_{\leftarrow}$) migration are also reported.}
	\label{fig:SMO}
\end{figure}

Here we report an additional mechanism for local ferromagnetism resulting from oxygen vacancies in an AFM material at finite temperature. We investigate the perovskite manganite \ce{SrMnO3} (SMO, see SI Fig. S1a) via density functional theory (DFT) calculations based on PBEsol+\textit{U}~\cite{perdew2008pbesol, Anisimov1991band, Dudarev1998electron} and performed with the \textsc{Quantum ESPRESSO}~\cite{giannozzi2009quantum, Giannozzi2017} package (see SI Section~S1 for additional computational details). In Fig.~\ref{fig:SMO}, we show the energy profile of oxygen-vacancy diffusive jumps from one of the energetically stable in-plane oxygen-vacancy sites, which we label C, to various neighboring oxygen-vacancy sites (see SI Table~S1 for formation energies - all raw data are also available on the Materials Cloud archive~\footnote{Materials Cloud URL to be added}). Also shown are the magnetic moments per cell at each intermediate image and the evolution of the oxidation state of the Mn sites that deviate from the normal \ce{Mn^{4+}} at some point along the path. Oxidation states computed based on \textit{d}-orbital occupations according to Ref.~\onlinecite{Sit2011} provide a more easily interpretable measure of charge localization compared to local magnetic moments. We observe that the cells acquire a sizable magnetic moment in the center portion of all migration paths, while magnetic moments at either of the end-points remain around zero as expected for an AFM ordered material. Magnetic moments at the saddle points are generally of magnitude 3.92~$\mu_\textrm{B}$/cell. Slightly smaller moments are observed for paths $\mathrm{C\rightarrow A_1}$ and $\mathrm{C\rightarrow A_4}$, which can be associated with their emerging metallicity and thus less localized excess charge (see SI Table~S2). For all paths we observe oxygen-vacancy migration barriers between 0.64 and 0.66~eV, in agreement with previous reports for SMO~\cite{Ricca:2021} and similar perovskite oxides~\cite{Cuong:2007, Walsh:2011, Klyukin:2017}.

To understand the emergence of a magnetic moment, we take the path $\mathrm{C}\rightarrow \mathrm{A}_1$ in Fig.~\ref{fig:SMO}a as an example. In the initial position, when the \ce{V_O} is at site C, its excess electrons are localized on \ce{Mn_1} and \ce{Mn_4} as shown by their respective \ce{Mn^{3+}}oxidation states. During diffusion, the O atom at site $\mathrm{A}_1$ moves towards the \ce{V_O} at site C, rotating around \ce{Mn_1} that temporarily becomes \ce{Mn^{5+}}, while the excess charge is localized on \ce{Mn_3} and \ce{Mn_4}. We note that the \ce{Mn^{5+}} oxidation state and the apparent violation of charge neutrality is due to the emergence of metallicity at the center of the path and a concomitant partial delocalization of the ``missing'' electron. In G-AFM SMO, \ce{Mn_3} and \ce{Mn_4} show the same spin polarization, thus resulting in a net magnetic moment for this excess charge accommodation, which is maximal at the saddle point. In the final configuration with a \ce{V_O} at site $\mathrm{A}_1$, the excess electrons are localized on \ce{Mn_1} and \ce{Mn_3}, which are adjacent to the vacancy and have opposite spin polarization, resulting in no net moment. Another observation from Fig. \ref{fig:SMO} is that the cell develops a positive moment for half of the paths, while it is negative for the other half. This is due to half the sites that a diffusing oxygen rotates around being up and the other half being down polarized.

\begin{figure}
	\includegraphics[width=0.5\columnwidth]{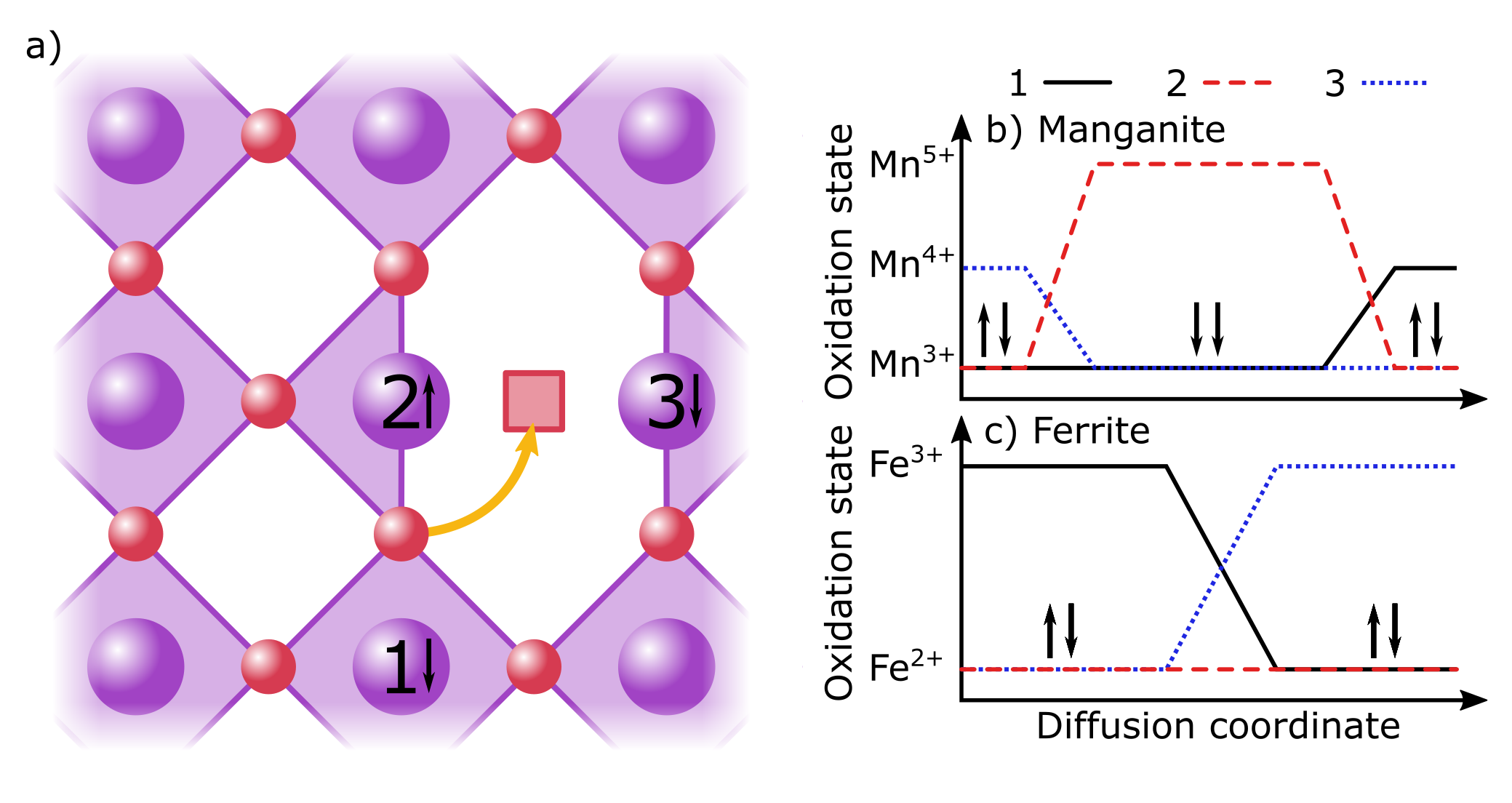}
	\caption{Schematic showing the difference in excess charge accommodation during a diffusive hop in G-type AFM materials: a) Diffusion path along with labels and magnetic polarization for the involved B-sites, b) and c) evolution of the oxidation states along the diffusion path in a manganite and ferrite respectively. The paired arrows show where the excess electron spins are aligned antiparallel or parallel.}
	\label{fig:schematic}
\end{figure}

This is schematically shown in Fig.~\ref{fig:schematic}a and b, where for a diffusive jump we observe oxidation-state changes on the surrounding sites 1, 2 and 3, and emergence of a local ferromagnetic moment in the center portion of the path. Similar changes in oxidation state were reported for \ce{LaFeO3}~\cite{Ritzmann:2013kn} and \ce{SrCoO3}~\cite{Tahini:2016formation}, where, however, a single charge transfer occurs from site 1 to site 3, without involving site 2. Thus, since sites 1 and 3 have the same spin polarization in an AFM material, no net moment emerges along the pathway as shown in Fig.~\ref{fig:schematic}c. It is important to note that in SMO the diffusing oxygen atom does not have to reach the saddle point for a net moment to emerge. Vibrations of oxygen atoms around their minimum-energy position and towards a \ce{V_O} will already induce net magnetic moments. The magnitude of these induced moments will increase as the temperature and thus the vibration amplitude increases.

The emergence of magnetic moments due to oxygen displacements in the vicinity of a \ce{V_O} seems exciting as it would lead to ferrimagnetic properties. We note, however, that displacements of the oxygen vacancy are equally likely in either direction (e.g. paths $\mathrm{C}\rightarrow \mathrm{A}_1$ and $\mathrm{C}\rightarrow \mathrm{A}_3$ have almost the same barriers but result in opposite moments), leading to local moments that will on average cancel. Breaking the symmetry to render these paths with opposite migration direction inequivalent is thus a prerequisite for the emergence of net moments and ferrimagnetism.

We initially attempted symmetry breaking by inducing a polar ferroelectric distortion via tensile epitaxial strain in SMO~\cite{Lee2010epitaxial, Becher2015strain}, which, however, resulted in metallicity along the entire path and therefore delocalized excess charge without clear changes in oxidation state. Another route for symmetry breaking is via heterostructuring, for example by replacing every second \ce{SrO} layer by a \ce{CaO} layer, yielding \ce{SrCaMn2O6} (SCMO, see SI Fig.~S1b). In this case, the formation energies for oxygen vacancies are lowest in the \ce{CaO} layer and largest in the \ce{SrO} layer (see SI Section~S3 for details). While the symmetry would have been broken for a \ce{V_O} in the \ce{MnO2} layer, for a \ce{V_O} in the \ce{CaO} layer, migration paths with opposite direction along $b$ have equivalent barriers but lead to opposite magnetic moments, resulting in no net moment.

\begin{figure}
	\includegraphics[height=0.8\textheight]{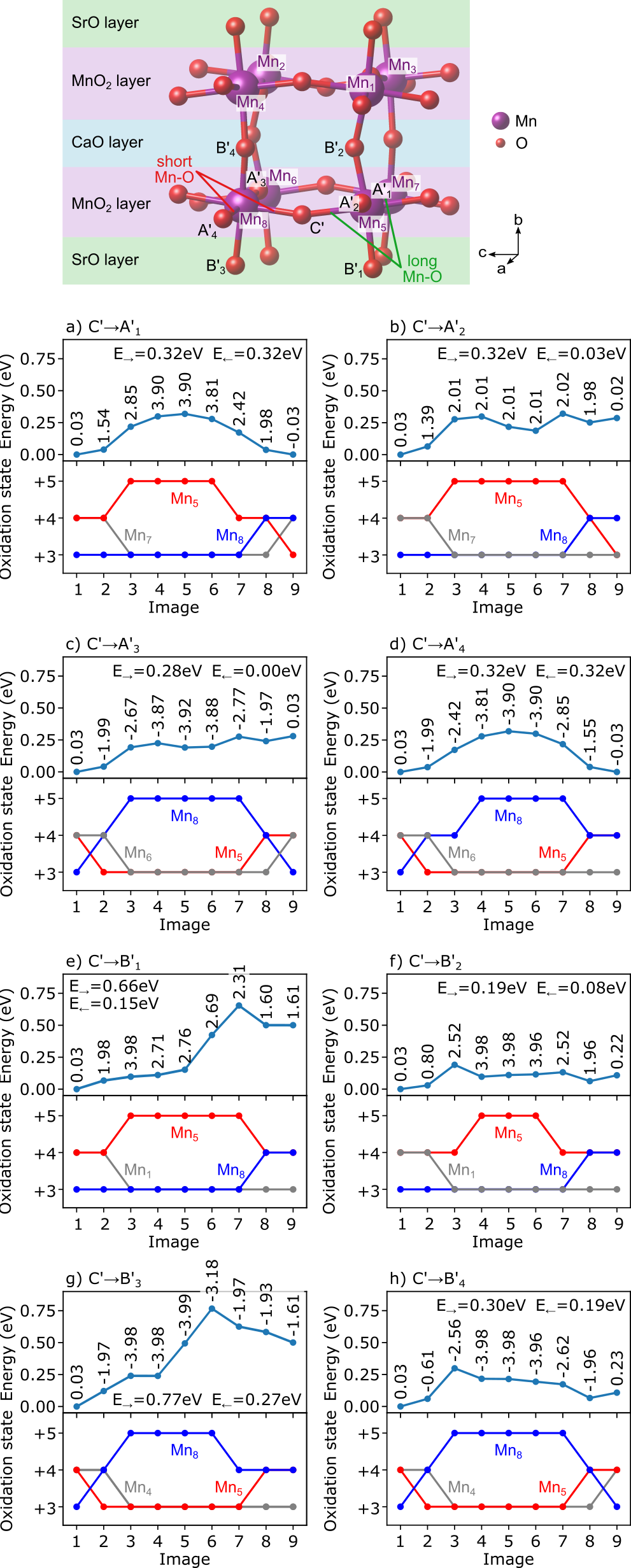}
	\caption{Energy profile for the migration of a \ce{V_O} in 4\% tensile strained SCMO from site $\mathrm{C}'$ to different neighboring sites shown in in the top panel (Sr and Ca atoms are omitted for clarity). For each image along the migration paths the total magnetization is reported in $\mu_\textrm{B}$/cell. Barriers for forward (E$_{\rightarrow}$) and reverse (E$_{\leftarrow}$) migration are also reported.}
	\label{fig:SCMO}
\end{figure}

Only when additionally applying tensile epitaxial strain to the SCMO heterostructure (SI Fig.~S1c), the desired symmetry breaking is obtained. Biaxial tensile strain applied in the $ac$ plane does not only shrink the $b$ lattice parameter, increase the average Mn--O bond length and alter octahedral rotations, but more importantly stabilizes a ferroelectric state with polarization in the $ac$ plane. This distortion results in alternating short and long Mn--O bonds in the $ac$-plane~\cite{Ricca2019,Ricca:2021}, as schematically shown at the top of Fig.~\ref{fig:SCMO}). Importantly, replacement of half of the Sr by Ca also opens the band gap and counteracts, at least in one spin channel, the metallicity observed in oxygen-deficient ferroelectric \ce{SrMnO3} (see SI section~S4).

\begin{table}
	\caption{\label{tbl:SCMO}Formation energies for oxygen vacancies in 4\% tensile strained SCMO at the sites shown in Fig.~\ref{fig:SCMO}.}
	\begin{tabular}{l|rr}
		\hline\hline
		Sites & Formation energy (eV) \\
 		\hline
 		$\mathrm{A}_1'$, $\mathrm{A}_4'$ & 1.524 \\
		$\mathrm{A}_2'$, $\mathrm{A}_3'$ & 1.804 \\
		$\mathrm{B}_1'$, $\mathrm{B}_3'$ & 2.025 \\
		$\mathrm{B}_2'$, $\mathrm{B}_4'$ & 1.632 \\
		$\mathrm{C}'$  & 1.524 \\
		\hline\hline 
	\end{tabular}
\end{table}

As shown in Table~\ref{tbl:SCMO} and expected due to chemical expansion~\cite{Adler2001chemical}, strain reduces the \ce{V_O} formation energies for all in-plane ($\mathrm{A}'$ and $\mathrm{C}'$) sites, compared to SMO and unstrained SCMO~\cite{Aschauer:2013,Ricca2019}. The large difference between \ce{V_O} formation energies on different $\mathrm{A}_i'$ and $\mathrm{B}_i'$ sites stems from heterostructuring. For the $\mathrm{A}_i'$ sites we observe that those, which, due to octahedral rotations, are closer to the SrO layer have a lower formation energy, which can be rationalized with the relative expanded volume in the SrO layer~\cite{Adler2001chemical}. For $\mathrm{B}_i'$ sites within the actual SrO and CaO layers, we predict significantly lower formation energies in the CaO layer compared to the SrO layer. Among the most stable in-plane sites we select the one labeled $\mathrm{C}'$ as the starting point for nudged elastic band calculations, but note that sites $\mathrm{A}_1'$ and $\mathrm{A}_4'$ that are equivalent in energy would lead to the same results.

The data in Fig.~\ref{fig:SCMO} shows that also in strained SCMO \ce{V_O} displacements induce local magnetic moments, the sign of which is dictated by the spin polarization of the Mn that the diffusing O atom rotates around. The Mn oxidation states also show a similar, although less regular evolution than in SMO, which can be linked with band-gap variations and metallicity towards the end points of the paths (see SI Fig.~S3). For in-plane migration (Fig.~\ref{fig:SCMO}a-d) we still observe paths with similar barriers but opposite magnetic moment. The-out-of-plane paths, however, are clearly affected by heterostructuring and the polar distortion. Pathways $\mathrm{C}'\rightarrow \mathrm{B}_2'$ and $\mathrm{C}'\rightarrow \mathrm{B}_4'$, when the \ce{V_O} displaces towards the CaO layer, have significantly lower barriers than pathways $\mathrm{C}'\rightarrow \mathrm{B}_1'$ and $\mathrm{C}'\rightarrow \mathrm{B}_3'$. Moreover displacement of the \ce{V_O} in the direction of the longer Mn--O bond is more favorable, leading to a very small barrier for pathway $\mathrm{C}'\rightarrow \mathrm{B}_2'$ that is, in fact, smaller than the one for the in-plane pathways. This imbalance of \ce{V_O} displacement energies will lead to more frequent displacements along the path $\mathrm{C}'\rightarrow \mathrm{B}_2'$ compared to all the other paths and hence a preferential rotation around \ce{Mn_5} with a net positive induced magnetic moment. For a \ce{V_O} on the energetically equivalent $\mathrm{A}_1'$ and $\mathrm{A}_4'$ sites, preferential displacements would lead to rotations around \ce{Mn_5} and \ce{Mn_6} respectively, also leading to net positive moments.

The combination of \ce{SrO-MnO2-CaO-MnO2} heterostructuring and a polar distortion induced by tensile epitaxial strain thus breaks the symmetry of atomic vibrations around oxygen vacancies in a way that induces net local moments with preferential spin polarization. These local moments lead to ferrimagnetism in the oxygen-deficient strained heterostructure.

An experimental realization of our prediction will require growing the \ce{SrO-MnO2-CaO-MnO2} heterostructure on a substrate that induces a suitable amount of tensile strain, such as \ce{SrTiO3} (3.55\% strain), at an oxygen partial pressure that leads to oxygen deficiency. Since manganites are known to equilibrate their oxygen content rather rapidly when in contact with air, capping the heterostructure will be crucial to preserve the oxygen deficiency~\cite{Chandrasena:2017dz}. The magnetic moments resulting from our predicted mechanism will show a fairly unusual temperature dependence. At low temperature, vibrations of oxygen atoms are minor, leading to very small induced moments, the antiferromagnetic property of the host material dominating. The induced moments will then increase with increasing temperature, leading to ferrimagnetic behavior of the oxygen deficient heterostructure. Once the temperature is sufficiently high to frequently overcome transition states, oxygen vacancies will disorder, leading again to, on average, canceling local moments and either antiferromagnetic (below the N\'eel temperature of the heterostructure) or paramagnetic behavior (above the N\'eel temperature of the heterostructure).

\begin{acknowledgments}
This research was supported by the NCCR MARVEL, a National Centre of Competence in Research, funded by the Swiss National Science Foundation (grant number 182892) and by the Swiss National Science Foundation Project 200021\_178791. Calculations were performed on UBELIX (http://www.id.unibe.ch/hpc), the HPC cluster at the University of Bern and on Piz Daint at the Swiss Supercomputing Center CSCS under project IDs s1033 and mr26.
\end{acknowledgments}

\bibliography{references}

%%%%%%%%%%%%%%%%%%%%%%%
% Begin SI
%%%%%%%%%%%%%%%%%%%%%%%

\clearpage
\clearpage
\renewcommand{\thepage}{S\arabic{page}}
\setcounter{page}{1}
\renewcommand{\thetable}{S\arabic{table}} 
\setcounter{table}{0}
\renewcommand{\thefigure}{S\arabic{figure}}
\setcounter{figure}{0}
\renewcommand{\thesection}{S\arabic{section}}
\setcounter{section}{0}
\renewcommand{\theequation}{S\arabic{equation}}
\setcounter{equation}{0}
\onecolumngrid

%create title
\begin{center}
\textbf{Supplementary information for\\\vspace{0.5 cm}
\large Ferrimagnetism induced by thermal vibrations in oxygen-deficient manganite heterostructures\\\vspace{0.3 cm}}
Moloud Kaviani,$^1$ Chiara Ricca,$^1$ and Ulrich Aschauer$^{1, 2, *}$

\small
$^1$\textit{Department of Chemistry and Biochemistry, University of Bern, Freiestrasse 3, CH-3012 Bern, Switzerland}

$^2$\textit{Department of Chemistry and Physics of Materials, University of Salzburg, Jakob-Haringer-Strasse 2A, 5020 Salzburg, Austria}

\vspace{0.3cm}
Email: ulrich.aschauer@plus.ac.at

\vspace{0.3cm}
(Dated: \today)
\end{center}

\section{Computational Details}\label{sec:comp_details}

All DFT calculations were performed with the {\sc{Quantum ESPRESSO}} package~\citeSI{SI_giannozzi2009quantum, SI_Giannozzi2017}, at the PBEsol+$U$~\citeSI{SI_perdew2008pbesol} level of theory. The Hubbard parameter for the Mn atoms was set to 4.26~eV, computed self-consistently for bulk \ce{SrMnO3}~\citeSI{SI_Ricca2019}. Ultrasoft pseudopotentials~\citeSI{SI_vanderbilt1990soft} with Ca($3s$, $3p$, $4s$), Sr($4s$, $4p$, $5s$), Mn($3p$, $4s$, $3d$), and O($2s$, $2p$) valence states were employed and wavefunctions expanded in plane waves with a kinetic-energy cut-off of 70~Ry and a cut-off of 840~Ry for the augmented density. A Gaussian smearing with a broadening parameter of 0.02 Ry was used in all cases.

Bulk \ce{SrMnO3} (SMO) was modeled as a 40-atom $2\times2\times2$ supercell of the 5-atom primitive cubic cell and included octahedral rotations leading to a $Pnma$ space group. The \ce{SrCaMnO3} (SCMO) heterostructure was built by substituting one plane of Sr atoms along the $b$-axis by Ca ions, which we find to be energetically preferred by 0.017 eV per 40-atom cell compared to ordering along the a and c axes. Biaxial epitaxial strain in the \textit{ac}-plane as imposed by a cubic substrate was simulated by adjusting the in-plane lattice parameters to equal length and at 90$^\circ$ with respect to each other~\citeSI{SI_Rondinelli:2011jk}. Reciprocal space for these 40-atom cells was integrated using a $6\times6\times6$ Monkhorst-Pack \textbf{k}-point grid. For stoichiometric cells, both lattice vectors and atomic position were relaxed, while for defective cells, only the atomic positions were relaxed, keeping the lattice vectors fixed at optimized values of the stoichiometric cell. In all cases, atomic forces were converged to within $5\times10^{-2}$~eV/\AA, while energies were converged to within $1.4\times10^{-5}$~eV.

A neutral oxygen vacancy (\ce{V_O}) was created in SMO and SCMO by removing one neutral oxygen atom. The \ce{V_O} formation energy ($E_\textrm{f}$) was computed according to Ref.~\citeSI{SI_freysoldt2014first}:
\begin{equation}
	E_\textrm{f} = E_\textrm{tot,def} - E_\textrm{tot,stoic} + \mu_\textrm{O} \\
\end{equation}
where $E_\textrm{tot,def}$ and $E_\textrm{tot,stoic}$ are the DFT total energies of the defective and stoichiometric cell, respectively, and $\mu_\textrm{O}$ is the O chemical potential, for which we assume the oxygen-rich limit ($\mu_\textrm{O} = \frac{1}{2}E(\textrm{O}_2)$ with $E(\textrm{O}_2)$ being the energy of an \ce{O2} molecule). Barriers for the \ce{V_O} migration were calculated using the climbing-image nudged elastic band (CI-NEB) method~\citeSI{SI_Henkelman2000}. Minimum energy pathways were relaxed until forces on each image converged below $1\times10^{-3}$~eV/\AA. Oxidation states were computed as described in Ref.~\citeSI{SI_Sit2011}.

\section{Bulk SMO}

As expected in highly symmetric bulk SMO, the \ce{V_O} formation energies are very similar for all sites. The sites in the $ac$ plane (perpendicular to the in-phase octahedral rotation axis), labeled A and C (see Fig. 1 in the main text for labels) being slightly more favorably by 25 meV than sites labeled B that lie along the in-phase rotation direction (see Table \ref{tbl:SMO}).

\begin{table}[h]
	\caption{\label{tbl:SMO}Formation energies for oxygen vacancies at the sites shown in Fig. 1 of the main text.}
	\begin{tabular}{l|r}
		\hline\hline
		Sites & Formation energy (eV)\\
 		\hline
		$\mathrm{A}_1$, $\mathrm{A}_2$, $\mathrm{A}_3$, $\mathrm{A}_4$ & 1.854 \\
		$\mathrm{B}_1$, $\mathrm{B}_2$, $\mathrm{B}_3$, $\mathrm{B}_4$ & 1.879 \\
		$\mathrm{C}$                                                   & 1.854 \\
		\hline\hline 
	\end{tabular}
\end{table}

While most migration pathways in bulk SMO have a maximum total magnetization of about 3.92~$\mu_\textrm{B}$/cell at the saddle point, we find a slightly smaller value of 2.68~$\mu_\textrm{B}$/cell for the pathways $\textrm{C}\rightarrow \textrm{A}_1$  (Fig.~1a in the main text) and $\textrm{C}\rightarrow \textrm{A}_4$ (Fig.~1d in the main text). As shown by the band gaps reported in Table~\ref{tbl:SMO_gap}, the final structures of these pathways have a more metallic character with a smaller band gap, resulting in a less clear excess charge localization on the involved \ce{Mn} atoms.

\begin{table}[hbt]
	\caption{\label{tbl:SMO_gap}Band gaps in the spin up and down channels as wells as the total (smaller among the two) gap.}
 	\begin{tabular}{l|rr|r}
 		\hline\hline
 		Site & Up (eV) & Down (eV) & Total (eV) \\
 		\hline
		$\mathrm{A}_1$ & 0.0063  & 0.0356    & 0.0044 \\
		$\mathrm{A}_2$ & 1.0488  & 0.0125    & 0.0125 \\
		$\mathrm{A}_3$ & 0.0179  & 1.0475    & 0.0179 \\
		$\mathrm{A}_4$ & 0.0351  & 0.0066    & 0.0015 \\
		\hline\hline 
  \end{tabular}
\end{table}

\FloatBarrier
\section{Unstrained \ce{SCMO}}\label{sec:si_scmo}

\begin{figure}[h!]
	\includegraphics[width=0.5\columnwidth]{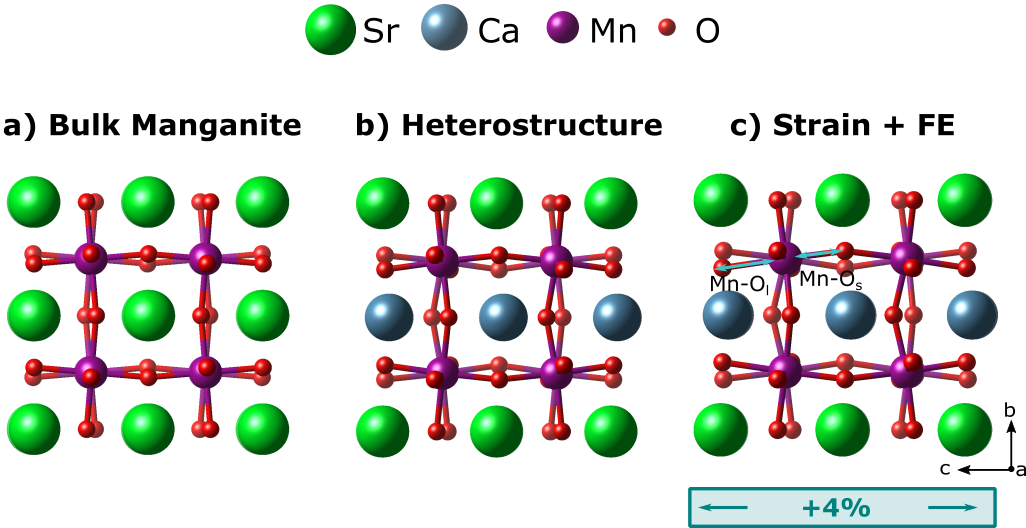}
	\caption{Structures of a) unstrained bulk SMO, b) unstrained SCMO and c) 4\% tensile strained SCMO. The long and short Mn--O bonds resulting from strain-induced ferroelectricity are indicated as Mn--O$_l$ and Mn--O$_s$ respectively.}
	\label{fig:structures}
\end{figure}

Heterostructuring could be an efficient method to break the symmetry around the O site, in order to favor the displacement of a \ce{V_O} along a specific direction and thus the emergence of a net moment. For this reason, we investigate oxygen vacancy formation and migration in a Sr-Ca manganite heterostructure with alternating Sr and Ca A-site planes along the $b$-axis as shown in Fig.~\ref{fig:structures}b).

\begin{figure}[h!]
	\includegraphics[width=\columnwidth]{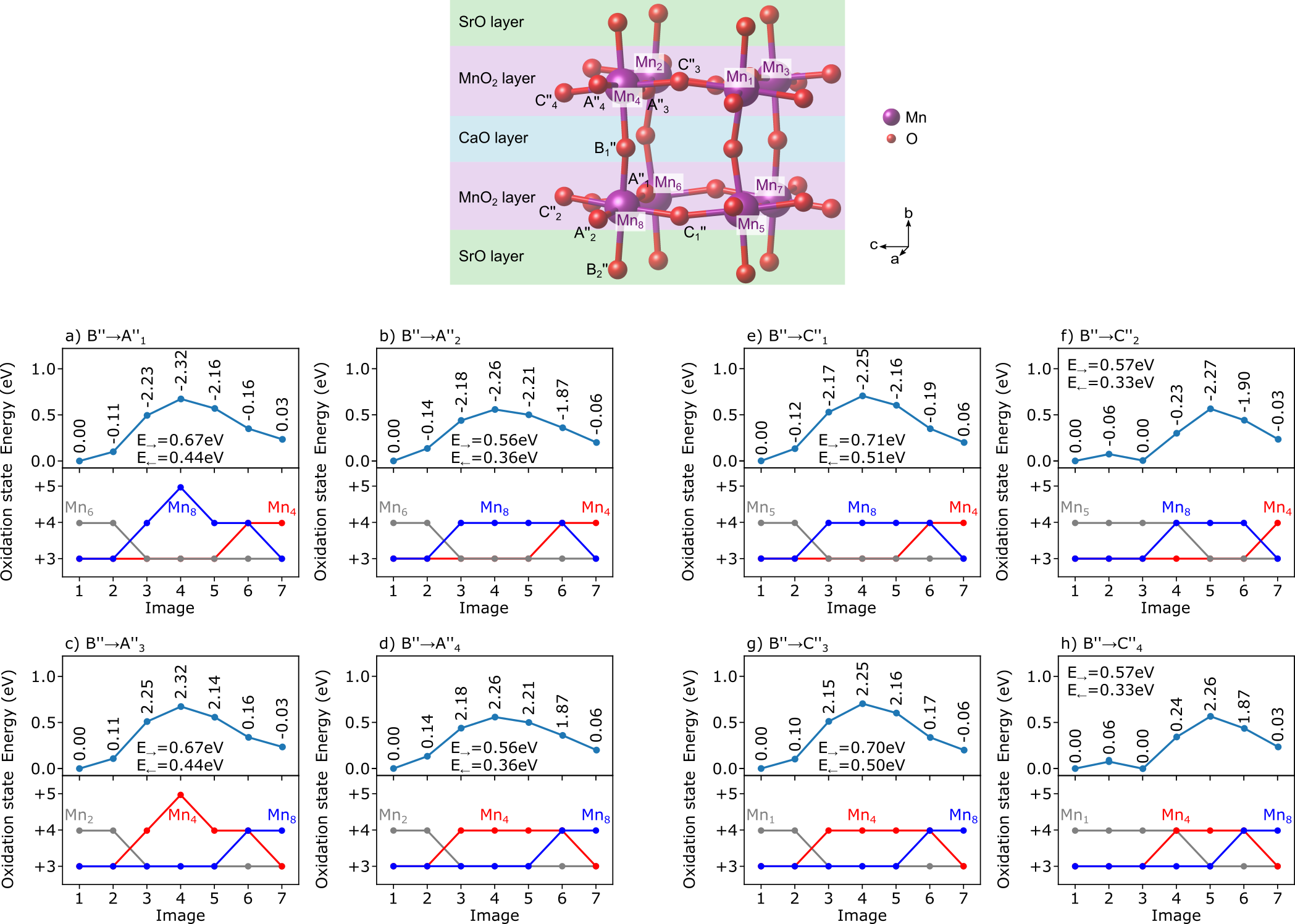}
	\caption{Barrier for the migration of one \ce{V_O} in unstrained SCMO from the most stable (out-of-plane) site $\mathrm{B}''$ to different neighboring sites (see top panel for site labels). For each image along the migration path the total cell magnetization is reported in $\mu_\textrm{B}$/cell. Barriers for forward (E$_{\rightarrow}$) and reverse (E$_{\leftarrow}$) migration are also reported.}
	\label{fig:SCMO_unstrained}
\end{figure}

The desired symmetry breaking is visible in the computed defect formation energies ($E_\textrm{f}$) for \ce{V_O} at different positions in the heterostructure. The most favorable \ce{V_O} position is a defect formed by removing one O atom from the Ca layer (configuration $\mathrm{B}''_1$ in Fig.~\ref{fig:SCMO_unstrained} and Table~\ref{tbl:SCMO_energies}). Therefore, while in SMO, the sites in the $ac$-plane were favored by a small energy difference (0.025 eV), now the sites perpendicular to this plane (along the b-axis) are favored by a larger energy difference of 0.2 eV. Removing an O along the same axis, but from the Sr layer, on the other hand, correspond to the most unfavorable defect configuration (see configuration $\mathrm{B}''_2$ in Fig.~\ref{fig:SCMO_unstrained} and Table~\ref{tbl:SCMO_energies}). For the defects in the $ac$-plane ($\mathrm{A}''$ and $\mathrm{C}''$ sites) created by the removal of one O atom in the \ce{MnO2} layer, we obtain similar formation energies, slightly larger for the O positions ($\mathrm{A}''_1$, $\mathrm{A}''_3$, $\mathrm{C}''_2$, $\mathrm{C}''_4$) which, due to the octahedral rotations, are further away from the Sr but closer to the Ca layer (see Table~\ref{tbl:SCMO_energies}) than the sites $\mathrm{A}''_2$, $\mathrm{A}''_4$, $\mathrm{C}''_1$, $\mathrm{C}''_3$.
\begin{table}[h!]
\caption{Formation energies ($E_\mathrm{f}$) for different \ce{V_O} sites in the unstrained SCMO heterostructure, as well as the distance of each considered O site from the Ca ($d_\mathrm{O-Ca}$) and Sr ($d_\mathrm{O-Ca}$) cations. See top panel of Fig.~\ref{fig:SCMO_unstrained} for the O site labels.
}
\begin{tabular}{l|r|rr}
\hline 
\hline 
Site & $E_\mathrm{f}$ (eV) & $d_\mathrm{O-Ca}$ (\AA) & $d_\mathrm{O-Sr}$ (\AA) \\
\hline 
$\mathrm{B}''_1$                                                       & 1.70 & 2.32-2.96 & 4.40-4.61 \\
$\mathrm{A}''_2$, $\mathrm{A}''_4$, $\mathrm{C}''_1$, $\mathrm{C}''_3$ & 1.90 & 2.56-3.07 & 2.42-2.65 \\
$\mathrm{A}''_1$, $\mathrm{A}''_3$, $\mathrm{C}''_2$, $\mathrm{C}''_4$ & 1.93 & 2.37-2.54 & 2.69-3.13 \\
$\mathrm{B}''_2$                                                       & 2.34 & 4.73-4.80 & 2.56-2.91 \\
\hline 
\hline 
\end{tabular}
\label{tbl:SCMO_energies}
\end{table}

O diffusion in presence of a \ce{V_O} takes place according to the same mechanism described in the main text for bulk \ce{SrMnO3}: during the migration, and in particular around the transition state, the Mn atom around which the migrating O rotates becomes \ce{Mn^{4+}}, while the remaining Mn atoms adjacent to the initial and final defect position accommodate the two extra electrons left in the lattice by the vacancy being reduced to \ce{Mn^{3+}}, as shown by the evolution of the oxidation state for selected Mn atoms for the considered migration paths in Fig.~\ref{fig:SCMO_unstrained}.

 As discussed in the main text for SMO, the reduced \ce{Mn^{3+}} in the central portion of the path have the same spin polarization, resulting in a net total magnetization of about 2.3~$\mu_\textrm{B}$/cell at the transition state of each path, slightly smaller that the one observed in bulk SMO. The computed O diffusion barriers are of the same order as the ones observed in SMO (0.56-0.71~eV). Interestingly, not all the paths are equivalent as in SMO, with the lower barriers observed for some of the paths.
 
 Unfortunately, despite this anisotropy of the migration barriers, paths which would result in opposite local magnetic moments show the same barrier. The symmetry breaking induced by the heterostructure is thus not sufficient for aligned local magnetic moments due to vacancy migration.

\FloatBarrier
\clearpage
\section{Band gaps}\label{sec:si_scmo_bandgap}

\begin{figure}[h!]
	\includegraphics[width=0.9\columnwidth]{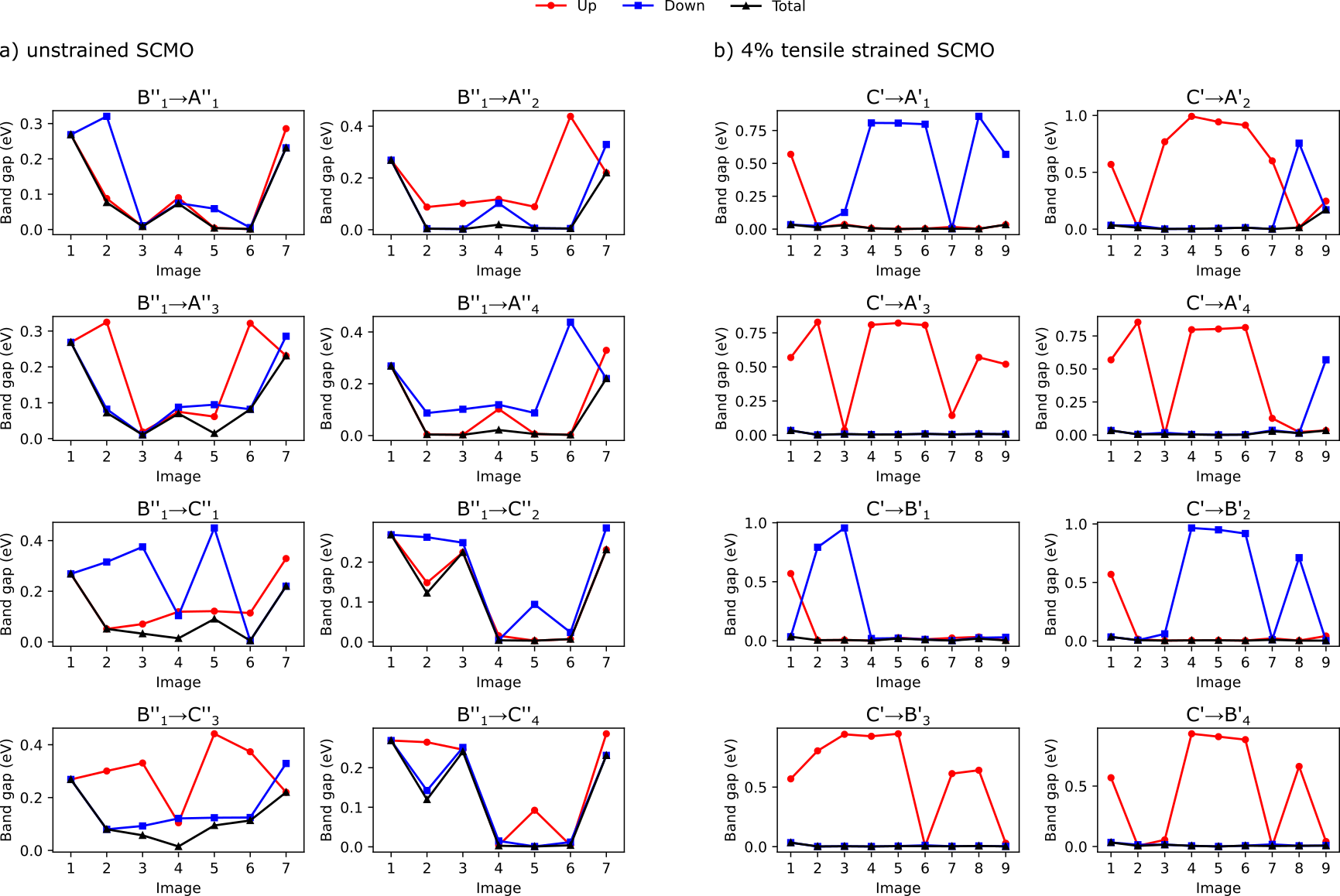}
	\caption{Evolution of the band gap during the migration of one \ce{V_O} in a) unstrained and b) 4\% tensile strained \ce{SrCaMnO3} along the considered diffusion paths.}
	\label{fig:SCMO_gaps}
\end{figure}

As shown in Fig. \ref{fig:SCMO_gaps}, Ca substitution and 4\% tensile strain lead to sizable gaps (0.7-1.0 eV) along a large portion of most migration paths, whereas just Ca substitution leads to small gaps, especially in the central portion of the path.

\bibliographystyleSI{apsrev4-1}
\bibliographySI{references}

\end{document}